\documentclass[12pt,authoryear]{article}
\usepackage{,amsmath,colonequals,color}
\usepackage{graphicx,natbib}
\usepackage{apalike}

\usepackage{amsfonts,marvosym}
\usepackage{algorithm}
\usepackage{bm}
\usepackage{geometry}
\usepackage{lscape}
\usepackage{hyperref}
\usepackage{mathtools}
\usepackage{caption}
\usepackage{theorem}
\usepackage{placeins}
\usepackage{enumerate}
\usepackage{tabulary}
\usepackage{listings}
\usepackage[usenames,dvipsnames,svgnames,table]{xcolor}

\newcommand{\be}{\begin{equation}}
\newcommand{\ee}{\end{equation}} 
\newcommand{\beq}{\begin{equation*}}
\newcommand{\eeq}{\end{equation*}}

\newcommand{\diag}{\,\mbox{diag}}

\newcommand{\ident}{\mathbf{I}}
\newcommand{\vecA}{\mathbf{A}}

\newcommand{\vecD}{\mathbf{D}}
\newcommand{\vecI}{\mathbf{I}}

\newcommand{\vecx}{\mathbf{x}}

\newcommand{\vecPsi}{\mbox{\boldmath$\Psi$}}

\newcommand{\varthet}{\mbox{\boldmath$\vartheta$}}
\newcommand{\vectheta}{\mbox{\boldmath$\theta$}}
\newcommand{\vecDelta}{\mbox{\boldmath$\Delta$}}
\newcommand{\del}{\mbox{\boldmath$\Delta$}}
\newcommand{\Esub}{\mathbb{E}}
\newcommand{\real}{\mathbb{R}}
\newcommand{\vecmu}{\mbox{\boldmath$\mu$}}

\newcommand{\mSigma}{\mbox{\boldmath$\Sigma$}}
\newcommand{\matsig}{\mSigma}
\newcommand{\vecLambda}{\mbox{\boldmath$\Lambda$}}

\newcommand{\zig}{\hat{z}_{ig}}

\theorembodyfont{\upshape}

\begin{document}

\title{Subspace Clustering with the Multivariate-$t$ Distribution}

\author{Angelina\ Pesevski$^*$, Brian\ C. Franczak$^{**}$ and Paul\ D. McNicholas$^*$\\[+2mm]
$^*${\small Department of Mathematics \& Statistics, McMaster University, Ontario, Canada.}\\
$^{**}${\small Department of Mathematics \& Statistics, MacEwan University, Alberta, Canada.}}

\date{}

\maketitle

\abstract{Clustering procedures suitable for the analysis of very high-dimensional data are needed for many modern data sets. In model-based clustering, a method called high-dimensional data clustering (HDDC) uses a family of Gaussian mixture models for clustering. HDDC is based on the idea that high-dimensional data usually exists in lower-dimensional subspaces; as such, an intrinsic dimension for each sub-population of the observed data can be estimated and cluster analysis can be performed in this lower-dimensional subspace. As a result, only a fraction of the total number of parameters need to be estimated and a computationally efficient parameter estimation scheme based on the EM algorithm was developed. This family of models has gained attention due to its superior classification performance compared to other families of mixture models; however, it still suffers from the usual limitations of Gaussian mixture model-based approaches. In this paper, a robust analogue of the HDDC approach is proposed. This approach, which extends the HDDC procedure to include the mulitvariate-$t$ distribution, encompasses 28 models that rectify the aforementioned shortcomings of the HDDC procedure. Our $t$HDDC procedure is fitted to both simulated and real data sets and is compared to the HDDC procedure using an image reconstruction problem that arose from satellite imagery of Mars' surface.}

\pagenumbering{arabic}

\section{Introduction}\label{sec:introduction}
\setcounter{page}{1}

Cluster analysis refers to the practice of using statistical approaches to detect subgroups within a given data set. These subgroups can represent a physical attribute not described by the given explanatory variables, e.g., gender, income tax bracket or blood type, which can reveal important relationships among the observed data and may be a crucial component in the effective analysis of a given data set.
Due to their construction, finite mixture models are very useful when modelling data that contain a finite collection of sub-populations because each component of the model can be used to represent one of these sub-populations. Reviews of the application of finite mixture models for clustering
are given by \citet{fraley2002model,bouveyron14} and \cite{mcnicholas16b}, and extensive details can be found in the monographs by \cite{mclachlan00} and \cite{mcnicholas16a}. 

The density of a parametric finite mixture distribution is 
\begin{equation}\label{eq:fmm}
f(\vecx\, | \, \varthet)=\sum_{g=1}^{G} \pi_g\, p_g (\vecx\, |\, \vectheta_g),
\end{equation}
where $\pi_g>0$, such that $\sum_{g=1}^{G} \pi_g=1$, are called mixing proportions, $p_g (\vecx\, |\, \vectheta_g)$ are the component densities and $\varthet=(\pi_1,\ldots,\pi_G,\vectheta_1,\ldots,\vectheta_G)$ is a vector containing the model parameters. Herein, we follow convention and refer to the application of finite mixture models for clustering as model-based clustering.

The general mixture given in~\eqref{eq:fmm} can be specified to contain components of any univariate or multivariate probability distribution. 
Until the last decade or so, the majority of work on model-based clustering using multivariate component densities focused on the Gaussian mixture model.
One of the first notable departures from Gaussianity was provided by \cite{peel00}, who utilized mixtures of multivariate-$t$ distributions for clustering. Despite rectifying a well known short-coming of the Gaussian mixture model by formulating a model that is robust to outliers, mixtures of multivariate-$t$ distributions have only gained popularity in the last few years \citep[see][for examples]{mclachlan2007extension,andrews2011extending,andrews2011extendingb,baek11,steane12, andrews2012model,lin14}. As the name suggests, mixtures of multivariate-$t$ distributions assume that each sub-population of the observed data follow the multivariate-$t$ distribution. As such, the density of a mixture of multivariate-$t$ distribution is formulated by writing the component density in~\eqref{eq:fmm} as
\begin{align} \label{eq:dens_mt}
p_g\left(\vecx\mid\vectheta_g\right) = f_t(\vecx\mid\vecmu_g,\mSigma_g,\nu_g) = \frac{ \Gamma\left[\left(\nu_g + p\right)/2\right]\lvert\mSigma_g\rvert^{-1/2}(\pi\nu_g)^{-p/2} }
{ \Gamma\left[\nu_g/2\right] \left[1 + \delta\left(\vecx,\vecmu_g\mid\mSigma_g\right)/\nu_g\right]^{\left(\nu_g+p\right)/2} },
\end{align}
where $\Gamma\left(\cdot\right)$ is the Gamma function, $p$ is the number of dimensions in the observed data set, $\vecmu_g$ is the component location parameter, $\mSigma_g$ is the component covariance matrix, $\nu_g$ parameterizes the degrees of freedom in each component, and $\delta(\vecx,\vecmu_g\mid\mSigma_g)=\left(\vecx - \vecmu_g\right)' \mSigma_g^{-1}\left(\vecx-\vecmu_g\right)$ is the Mahalanobis distance between $\vecx$ and $\vecmu_g$ for $g=1,\ldots,G$. 

A family of mixture models emerges when we introduce constraints on the component densities. Some families of Gaussian mixture models are well established and widely used, e.g., the Gaussian parsimonious clustering models \citep[GPCM;][]{celeux1995gaussian} which arise from constraints being imposed on the eigen-decomposed covariance structure in a Gaussian mixture model. This eigen-decomposition is
$\mSigma_g=\lambda_g \vecD_g \vecA_g\vecD_g^{'},$
where $\lambda_g$ is a constant, $\vecD_g$ is a matrix of eigenvectors, and $\vecA_g$ is a diagonal matrix with $|\vecA_g|=1$ and entries proportional to the eigenvalues of $\mSigma_g$. Applying a combination of the constraints: $\lambda_g=\lambda, \vecA_g=\vecA, \vecD_g=\vecD, \vecD_g=\vecI$, $\vecA_g=\vecI$, where $\vecI$ is the identity matrix of the appropriate dimension, across the groups in the data creates a family of fourteen models and allows for various shapes and sizes of clusters (Table~\ref{tab:mclust2}). In more than half of these fourteen models there are $\mathcal{O}(p^2)$ free parameters to be estimated; hence, with higher dimensions, it can be very computationally inefficient to use these models. The GPCMs are supported by the \textsf{R} packages \texttt{mclust} \citep{fraley2002model} and \texttt{mixture} \citep{mixture2,mixture1}. 

\begin{table}[!ht]
\centering\caption{Nomenclature, covariance structure, and number of free covariance parameters for each member of the GPCM family.}\label{tab:mclust2}
{\small\begin{tabular*}{0.9\textwidth}{@{\extracolsep{\fill}}lllllr}
\hline
Model &  Volume & Shape & Orientation & $\matsig_g$ & Free covariance parameters\\
\hline
EII & Equal & Spherical & --& $\lambda\ident$ & 1\\
VII & Variable & Spherical & -- & $\lambda_g\ident$ & $G$\\
EEI & Equal & Equal & Axis-Aligned & $\lambda\vecA$ & $p$\\
VEI & Variable & Equal & Axis-Aligned & $\lambda_g\vecA$ & $p+G-1$\\
EVI & Equal & Variable & Axis-Aligned & $\lambda\vecA_g$ & $pG-G+1$\\
VVI & Variable & Variable & Axis-Aligned & $\lambda_g\vecA_g$ & $pG$\\
EEE & Equal & Equal & Equal & $\lambda\vecD\vecA\vecD'$ & $p(p+1)/2$\\
EEV & Equal & Equal & Variable & $\lambda\vecD_g\vecA\vecD_g'$ & $Gp(p+1)/2 - (G-1)p$\\
VEV & Variable & Equal & Variable & $\lambda_g\vecD_g\vecA\vecD_g'$ & $Gp(p+1)/2 - (G-1)(p-1)$\\
VVV & Variable & Variable & Variable & $\lambda_g\vecD_g\vecA_g\vecD_g'$ & $Gp(p+1)/2$\\
EVE & Equal & Variable & Equal & $\lambda\vecD\vecA_g\vecD'$ & $p(p+1)/2 +(G-1)(p-1)$\\
VVE & Variable & Variable & Equal & $\lambda_g\vecD\vecA_g\vecD'$ & $p(p+1)/2 + (G-1)p$\\
VEE & Variable & Equal & Equal & $\lambda_g\vecD\vecA\vecD'$ & $p(p+1)/2 + (G-1)$\\
EVV & Equal & Variable & Variable & $\lambda\vecD_g\vecA_g\vecD_g'$ & $Gp(p+1)/2-(G-1)$\\
\hline
\end{tabular*}}
\end{table}

The multivariate-$t$ analog of the GPCM family for the mixtures of multivariate-$t$ distributions is the $t$EIGEN family \citep{andrews2012model}. These models use the same eigen-decomposition as the GPCM family and therefore 
the same constraints mentioned above can be applied, in addition to $\nu_g=\nu$. By combining these constraints, a total of 28 different models are derived. In \textsf{R}, all 28 models are supported by the \texttt{teigen} package \citep{andrews2014teigen,andrews17}.

Another popular family of mixture models are the Parsimonious Gaussian Mixture Models \citep[PGMM;][]{mcnicholas2008parsimonious}. These models are an extension of the mixture of factor analyzers \citep{ghahramani1996factor} whose component covariance matrices are written as $\mSigma_g=\vecLambda_g\vecLambda_g' + \vecPsi_g$, where $\vecLambda_g$ is a $p \times q$ loading matrix with $q < p$, and $\vecPsi_g$ is a diagonal $p\times p$ matrix with positive entries
for $g=1,\ldots,G$. By imposing constraints on $\vecLambda_g$ and $\vecPsi_g$ across the components, \cite{mcnicholas2008parsimonious} introduced eight parsimonious models in which the number of free parameters is $\mathcal{O}(p)$ so that the number of covariance parameters grows linearly with dimension. For this reason, these models are more appropriate than the GPCMs for high-dimensional data. These models can be implemented via the \texttt{pgmm} package for \textsf{R} \citep{mcnicholas2015pgmm}. Note: \cite{mcnicholas2010model} extended the PGMMs to include four new models by setting $\vecPsi_g = \omega_g\del_g$, where $\omega_g\in\mathbb{R}^+$ and $\del_g = \diag{\left\{\delta_1,\delta_2,\ldots,\delta_p \right\}}$ is a noise matrix, such that $\lvert\del_g\rvert = 1$.

The multivariate-$t$ analogue of the PGMMs, known as the mixture of multivariate $t$-factor analyzers (MM$t$FA) were introduced by \cite{mclachlan2007extension} and extended by \citet{andrews2011extending,andrews2011extendingb}. In the MM$t$FAs, the component covariance stucture is also parameterized as $\mSigma_g= \vecLambda_g \vecLambda_g' + \vecPsi_g$. By applying the constraints: $\vecPsi_g=\psi_g \vecI$, $\vecLambda_g=\vecLambda$, and $\nu_g=\nu$, \cite{andrews2011extending} created a family of six models, whose covariance parameters grow linearly with $p$, and \cite{andrews2011extendingb} extended this to a family of 24 models. 
It is worth noting that the probabilistic principal $t$-component analyzer model MPP$t$CA model is a special case of the MM$t$FA model, where $\vecPsi_g=\psi_g \vecI$. 
This family of 24 models is supported by the \texttt{mmtfa} package \citep{andrews2015mmtfa}.

\cite{bouveyron2007high} proposed a high-dimensional data clustering (HDDC) technique that is also based on an eigen-decomposition of the covariance structure of the Gaussian mixture model. This technique projects the data into a lower-dimensional subspace spanned by a subset of the eigenvectors of $\mSigma_g$. 
Formally, given a data set $\{\vecx_1,\ldots,\vecx_n \}$ of $n$ data points in $\real^p$ with $G$ sub-populations, this method assumes that high-dimensional data mostly rests in lower-dimensional subspaces. This assumption can drastically reduce the number of covariance parameters that require estimation and result in an efficient parameter estimation scheme. As with the GPCMs, \cite{bouveyron2007high} lets $\vecD_g$ be the orthogonal matrix of eigenvectors of $\mSigma_g$, but instead considers a block-diagonal matrix, $\vecDelta_g$, which contains the eigenvalues of $\mSigma_g$. Formally, $\vecDelta_g$ has the following form: 
\begin{equation}\label{eq:dg_form}
\vecDelta_g = \left( \begin{array}{ccc|ccccc}
a_{1g}& & 0 & & & & &\\ 
&\ddots & & & & \textbf{0} & &\\
0 & & a_{d_g g} & & & & &\\
\hline
& & & b_g & & & & 0 \\
& & & & \ddots & & &  \\
& \textbf{0} & & & & \ddots & &  \\
& & & & & &\ddots & \\
& & & 0  & & & & b_g \\
 \end{array} \right),
\end{equation}
where the upper left block is of size $d_g \times d_g$, where $d_g \in \{1,p-1 \}$ is the intrinsic dimension in each component, or cluster, and the lower right block is of size $(p-d_g) \times (p-d_g)$, with $a_{jg} > b_g$, for $j=1,\ldots,d_g$ for $g=1,\ldots,G$. \cite{bouveyron2007high} proposed two methods for estimating the intrinsic dimension in each component of this eigen-decomposed GMM. The first approach utilizes the scree-test of \cite{cattell1966scree} and
the second approaches utilizes the probabilistic Bayesian Information Criterion \citep[BIC;][]{schwarz1978estimating}, which is given by $\text{BIC}=2l(\hat{\varthet})-\rho \log n$, where $\rho$ is the number of parameters in the model, $n$ is the number of observations, and $l(\hat{\varthet})$ is the maximized log-likelihood value.
 The eigenvectors associated with the eigenvalues $a_{jg}$, for $j=1,\ldots,d_g$, span a subspace $\Esub_g \in \real^{d_g}$ for each cluster, such that $\vecmu_g \in \Esub_g$. The affine subspace $\Esub_g^{\perp}$ is defined such that $\Esub_g \otimes \Esub_g^{\perp}=\real^p$ and $\vecmu_g \in \Esub_g^{\perp}$. Each observation $\vecx_i$ is then projected onto the subspace $\Esub_g$, which is called the specific subspace of the $g${th} group since most of the data are assumed to live on or near this subspace. 
This decomposition leads to 28 possible models by constraining the parameters $[a_{jg}, b_g, \vecD_g, d_g]$ across the $G$ components. Of these 28 models, 14 have been implemented in the \textsf{R} package \texttt{HDclassif} \citep{hddccite}. 

Like the other Gaussian mixture model based approaches, this technique will suffer from the usual limitations, i.e., its parameter estimation scheme will not be robust to outliers. Herein, we discuss the derivation of a multivariate-$t$ analogue of the HDDC approach. This paper proceeds as follows: in Section~\ref{sec:method}, we outline the derivation of a multivariate-$t$ high dimensional data clustering (tHDDC) approach and present a computationally efficient parameter estimation scheme to fit the resulting models, in Section~\ref{sec:applic} we assess the classification performance of this novel family of models using a simulation study and three real data sets, and in Section~\ref{sec:discuss} we conclude with a discussion and suggestions for future work.

\section{Methodology} \label{sec:method}
We now lay out some groundwork for the newly proposed tHDDC approach. As previously mentioned, it is the $t$-analogue of HDDC method described in Section~\ref{sec:introduction}. As with all the families introduced earlier, the goal is to cluster a given data set $\{\vecx_1,\ldots,\vecx_n \}$ in $\real^p$ into $G$ homogeneous groups. The data are modelled by the general density in \eqref{eq:fmm}, with the multivariate component density given in \eqref{eq:dens_mt}. The general multivariate-$t$ mixture model requires the estimation of the full covariance structure, so the number of parameters to estimate is $\mathcal{O}(p^2)$. As \cite{bouveyron2007high} describe, via the \textit{empty space} phenomenon \citep{scott1983probability}, we can assume that most of the data live around lower-dimensional subspaces. By performing clustering in these lower-dimensional subspaces, the number of parameters to be estimated is reduced significantly.

\subsection{The General $t$HDDC model}\label{sec:general_thddc}

Analogous to the HDDC approach, we specify $\vecD_g$ to be the orthogonal matrix of eigenvectors and 
$\vecDelta_g=\vecD_g' \mSigma_g \vecD_g$,
where $\vecDelta_g$ is a class specific matrix of the form given in~\eqref{eq:dg_form}.
Following \cite{bouveyron2007high} we define $$P_g(\vecx)=\tilde{\vecD}_g \tilde{\vecD}_g' (\vecx-\vecmu_g) +\vecmu_g $$ as the projection of $\vecx$ on $\Esub_g$ and $$P_g^{\perp}(x)=\bar{\vecD}_g \bar{\vecD}_g' (\vecx-\vecmu_g) +\vecmu_g $$ as the projection of $\vecx$ on $\Esub_g^{\perp}$, where $\tilde{\vecD}_g$ consists of the first $d_g$ columns of $\vecD_g$, concatenated with $p-d_g$ zero columns and $\bar{\vecD}_g=\vecD_g-\tilde{\vecD}_g$. Each $t$HDDC model has parameters $a_{jg},b_g,\vecD_g,d_g,\nu_g$ for $j=1,\ldots,d_g$ and $g=1,\ldots,G$. Applying group-wide constraints to these parameters can lead to a total of 56 possible models. (The $t$HDDC analogues of the HDDC models available in \texttt{HDclassif} are listed in Table~\ref{tab:models}).

\begin{table}[!ht]
\caption{Nomenclature, covariance decomposition and number of free covariance parameters for the $t$HDDC models. For constaints on $a_{gi}$, U represents unconstrained, D represents constrained accross dimension, G represents constrained across groups and C represents constrained accross both dimension and group. For all other components, U and C are is unconstrained and constrained across groups, respectively. For the number of free parameters, $\rho=Gp+G+1$ is the number of parameters required to estimate the mean and proportions. The number of parameters required to estimate $\tilde{ D}_g$, $\tilde{D}$ and $s= \sum_{g=1}^G d_g$ are $\tau=d[p-(d+1)/2] $  and $\bar{\tau}=d_g[p-(d_g+1)/2] $ \label{tab:models}}
\centering
\begin{tabular*}{0.9\textwidth}{@{\extracolsep{\fill}}c c c c c c l}
\hline
Model & $a_{jg} = a_{g}/a_j$ & $b_g = b$ & $\vecD_g = \vecD$& $d_g = d$ & $\nu_g=\nu$ & Number of \\
& & & & & &Covariance Parameters  \\
\hline
\\[-2mm]
UUUUU &  U&  U&  U&  U & U& $\rho + \bar{\tau} +3G+s$ \\
UCUUU &  U&  C &  U&  U & U &$\rho + \bar{\tau} +2G+s+1$  \\
DUUUU &   D &  U&  U&  U& U & $\rho + \bar{\tau} +4G$ \\
CUUUU &   C &  U&  U &  U & U& $\rho + \bar{\tau} +3G+1$  \\
DCUUU & D & C &U & U &U &$\rho + \bar{\tau} +3G+1$ \\
CCUUU& C& C &U &U &U &$\rho + \bar{\tau} +2G+2$  \\
UUUCU & U &U &U & C & U &$\rho + G(\tau+d+2) +1$ \\
UCUCU &U &  C &  U &  C & U & $\rho + G(\tau+d+1) +2$ \\
DUUCU &D& U& U& C  &U& $\rho + G(\tau+2+1) +1$ \\
CUUCU &C & U  & U  & C &U& $\rho + G(\tau+2) +2$  \\
DCUCU& D& C& U  & C &U& $\rho + G(\tau+2) +2$  \\
CCUCU& C& C  & U & C &U& $\rho + G(\tau+1) +3$ \\
GCCCU & G& C  & C & C  &U& $\rho+\tau+d+G+2$\\
CCCCU & C& C  & C & C  &U& $\rho+\tau+G+3$\\
\\[-2mm]
UUUUC &  U&  U&  U&  U & C& $\rho + \bar{\tau} +2G+s+1$ \\
UCUUC &  U&  C &  U&  U & C &$\rho + \bar{\tau} +G+s+2$  \\
DUUUC &   D &  U&  U&  U& C & $\rho + \bar{\tau} +3G+1$ \\
CUUUC &   C &  U&  U &  U & C& $\rho + \bar{\tau} +2G+2$  \\
DCUUC & D & C &U & U &C &$\rho + \bar{\tau} +2G+2$ \\
CCUUC& C& C &U &U &C &$\rho + \bar{\tau} +G+3$  \\
UUUCC & U &U &U & C & C &$\rho + G(\tau+d+1) +2$ \\
UCUCC &U &  C &  U &  C & C & $\rho + G(\tau+d) +3$ \\
DUUCC &D& U& U& C  &C& $\rho + G(\tau+2) +2$ \\
CUUCC &C & U  & U  & C &C& $\rho + G(\tau+1) +3$  \\
DCUCC& D& C& U  & C &C& $\rho + G(\tau+1) +3$  \\
CCUCC& C& C  & U & C &C& $\rho + G\tau +4$ \\
GCCCC & G& C  & C & C  &C& $\rho+\tau+d+3$\\
CCCCC & C& C  & C & C  &C& $\rho+\tau+4$\\
\hline
\end{tabular*} 
\end{table} 

\subsection{The ECM for the General $t$HDDC Model}
In model-based clustering, the EM algorithm \citep{mclachlan08} is the usual choice for parameter estimation. It is an iterative procedure that alternates between two steps: an E-step and a M-step, until convergence is reached. On the E-step, the expected value of the complete-data log-likelihood is updated given the current estimates of the parameters. In the M-step, the same complete-data log-likelihood is maximized in terms of the model parameters. We use a variation of the EM algorithm called the expectation conditional-maximization (ECM) algorithm \citep{meng1993ECM}, which replaces each M-step with multiple CM-steps. For each $t$HDDC model the complete-data is made up of the observed $\vecx_i$, the latent $u_{ig}$, and the missing $z_{ig}$ for $i=1,\ldots,n$ and $g=1,\ldots,G$. Note that the $u_{ig}$ is a realization of a gamma distributed random variable, $U_{ig}$ that arises because we exploit the fact that the multivariate-$t$ distribution is a normal-variance mean mixture \citep{barndorff82,peel00}, whereas the $z_{ig}$ are introduced to represent component membership. Formally, we write that
\begin{equation}
z_{ig}=
\left\{
	\begin{array}{ll}
		1  & \mbox{if observation $\vecx_i$ belongs to component $g$ } \\
		0 & \mbox{otherwise. }
	\end{array}
\right.
\end{equation}
\subsubsection{The E-step}
For the general finite mixture model, the component indicator variables are usually replaced by their expected values,
\begin{equation*}\label{eq:zig}
\mathbb{E}[Z_{ig}\mid\vecx_i]=\frac{\pi_gp_g(\vecx\mid\varthet_g)}{\sum_{h=1}^G\pi_hp_h(\vecx\mid\varthet_h)}=:\zig,
\end{equation*}
for $i=1,\ldots,n$ and $g=1,\ldots,G$. Unfortunately, this usually requires the computation of both the determinant and inverse of a $p\times p$ covariance matrix. To avoid these potentially cumbersome calculations, we follow \cite{bouveyron2007high} and derive a cost function that utilizes the projection functions: $P_g(x)$ and $P_g^{\perp}(x)$, defined in Section~\ref{sec:general_thddc}. The derivation of the cost function is as follows: first, note that we can write
\begin{align*}
-2\log{f_t(\vecx\mid\vecmu , \mSigma ,\nu)  }
&= -2\log{ \Gamma\left[(\nu+p)/2\right]} + 2\log{\Gamma\left[\nu/2\right]} + p\left(\log{\nu} + \log{\pi}\right) \\
&+ \left(\nu+p\right)\log{\left[1 + \frac{1}{\nu}\left(\lvert\lvert\vecmu - P\left(\vecx\right)\rvert\rvert^2_{\vecA} + \frac{1}{b}\lvert\lvert\vecx - P'\left(\vecx\right)\rvert\rvert^2\right)  \right] } \\
&  + \sum_{j=1}^{d}{ \log{a_j} } + (p-d)\log{b},
\end{align*}
where $\lvert\lvert\vecx\rvert\rvert^2_{\vecA} = \vecx\vecA\vecx'$ with $\vecA = \tilde{\vecD}_g\vecDelta_g\tilde{\vecD}_g'$, and all other values are as previously defined. So, on the $E$-step of the proposed ECM algorithm we replace each $z_{ig}$ with
\begin{equation*}
\hat{z}_{ig} = \frac{1}{ \sum_{h=1}^G{ \exp{\left[ \frac{1}{2}\left(K_g\left(\vecx_i\right) - K_h\left(\vecx_i\right)\right) \right]} } },
\end{equation*}
where we refer to
\begin{align*}
K_g\left(\vecx_i\right) & = 
-2\log{ \Gamma\left[(\nu_g+p)/2\right] } + 2\log{\Gamma\left[\nu_g/2\right]} + {\color{black}(p-d_g)\log{b_g}} + p\left(\log{\nu_g} + \log{\pi}\right) \\
&~~~+ \left(\nu_g+p\right)\log{\left[1 + \frac{1}{\nu_g}\left({\color{black}\lvert\lvert\vecmu_g - P_g\left(\vecx\right)\rvert\rvert^2_{\vecA_g}} + {\color{black}\frac{1}{b_g}\lvert\lvert\vecx - P_g'\left(\vecx\right)\rvert\rvert^2}\right)  \right] } \\
&~~~ + {\color{black}\sum_{j=1}^{d_g}{ \log{a_{jg}} }} - 2\log{\pi_g},
\end{align*}
for $i=1,\ldots,n$ and $g=1,\ldots,G$, as the cost function. Each $u_{ig}$ is then replaced by their expected values
\begin{align*}\label{eq:uig}
\mathbb{E}[U_{ig}\mid\vecx_i,z_{ig}=1] &= \frac{\nu_{g}+p}{\nu_{g}+\lvert\lvert\vecmu_g - P\left(\vecx_i\right)\rvert\rvert^2_{\vecA} + \frac{1}{b_g}\lvert\lvert\vecx - P'\left(\vecx_i\right)\rvert\rvert^2} =:\hat{u}_{ig},\\
\end{align*}
for $i=1,\ldots,n$ and $g=1,\ldots,G$ \citep[cf.][]{peel00,andrews2012model}.

\subsubsection{The CM-steps}
On the first CM-step we update the mixing proportions and component location parameter using
\begin{equation*}
\hat{\pi}_g=\frac{n_g}{n}\quad\text{and}\quad\hat{\vecmu}_g = \frac{\sum_{i=1}^n \hat{z}_{ig} \hat{u}_{ig} \vecx_i}  {\sum_{i=1}^n \hat{z}_{ig} \hat{u}_{ig}},
\end{equation*}
respectively, where $n_g=\sum_{i=1}^n \hat{z}_{ig}$. The degrees of freedom parameter, ${\nu}_g$, is updated using the closed form approximation given in \cite{andrews17}. Formally, we let 
\begin{equation*}
\hat\nu_g \approx \frac{-\exp(k) + 2 \exp \left(  \varphi\left(\frac{\hat{\nu}_g^{\text{old}}}{2}\right)+\frac{1 - \hat{\nu}_g^{\text{old}}}{2}     \right) \exp(k) }{1-\exp(k)}
\end{equation*}
with
\begin{align*}
k = -1 &-\frac{1}{n_g} \sum_{g=1}^G \sum_{i=1}^n \hat{z}_{ig}\left(\log\hat{u}_{ig} -\hat{u}_{ig}\right) - \varphi\left(\frac{\hat{\nu}_g^{\text{old}}+p}{2}\right)+\log\left(\frac{\hat{\nu}_g^{\text{old}}+p}{2}\right)
\end{align*}
where $\hat{\nu}_g^{\text{old}}$ is the estimate of $\nu_g$ from the previous iteration of this ECM algorithm, and $\varphi\left(\cdot\right)$ is the digamma function.

%

For each tHDDC model, the updates on the second CM step are analogous to those given in \cite{bouveyron2007high}. For illustrative purposes, we outline how to update each covariance parameter for the UUUUU model, i.e., the model where $a_{jg},b_g,\vecD_g,d_g,\nu_g$ are free to vary across all $g=1,\ldots,G$ and $j=1,\ldots,d_g$. 

First, we calculate the intrinsic dimension, $d_g$. For each value of $j \in \left\{1,p-1\right\}$ we compute 
\begin{align}\label{eq:dg_ll}
l\left(\hat{\varthet}\right) &= -\frac{n}{2}\left( d_j\log{a_{jg}}+(p-d_j)\log{b_g} - 2\log{\pi_g} + \log{\nu_g} \right. \nonumber\\
&~~~\quad\left. + \log{\pi} - 2\log{\Gamma{\left[\left(\nu_g+p\right)/2\right]}} + 2\log{\Gamma\left(\nu_g/2\right)} \right)
\end{align}
and set $d_g$ equal to the value of $d_j$ that maximizes the BIC values associated with the log-likelihood values found using~\eqref{eq:dg_ll}. 
Then we let $\vecD_g$ be the eigenvectors of $\hat{\mSigma}_g$, where 
\begin{equation}
\hat{\mSigma}_g=\frac{1}{n_g} \sum_{i=1}^n \hat{z}_{ig} \hat{u}_{ig} (\vecx_i-\hat{\vecmu}_g) (\vecx_i-\hat{\vecmu}_g)'.
\end{equation}
Each $a_{jg}$, for $j=1,\ldots,d_g$ and $g=1,\ldots,G$ is then replaced with the first $d_g$ eigenvalues of $\hat{\mSigma}_g$ and we estimate $b_g$ using
\begin{equation}\label{eq:b1}
\hat{b}_g = \frac{1}{(p-d_g)}\left( \text{tr} \left(\hat{\mSigma}_g \right)- \sum_{j=1}^{d_g}a_{jg}\right).
\end{equation}

\subsection{Computational Considerations}
For the proposed ECM algorithm, we initialize each model using either k-means clustering or random starting values and use the Aitkin's acceleration \citep{aitken26} procedure to determine if the algorithm has converged. That is, we consider this ECM algorithm to have converged when $l_{\infty}^{(k+1)}-l^{(k)} < \epsilon$, where $\epsilon=10^{-2}$ \citep[see][]{lindsay1995mixture,mcnicholas10a}. In this criterion, $l^{(k)}$ is the log-likelihood value at iteration $(k)$ and $l_{\infty}^{(k+1)}$ is the asymptotic estimate of the log-likelihood at iteration $(k+1)$. Formally, 
\begin{equation}
l_{\infty}^{(k+1)}=l^{(k)} + \frac{1}{1-a^{(k)}}\left(l^{(k+1)}-l^{(k)}\right),
\end{equation}
where 
$$a^{(k)}=\left(l^{(k+1)}-l^{(k)}\right)/\left(l^{(k)}-l^{(k-1)}\right).$$ 

All analyses are performed in \textsf{R} version 3.3.2 \citep{Rcite} for Linux 6.5\footnote{\label{myfootnote}Using a 32-core Intel Xeon E5 server with 256GB RAM running 64-bit CentOS}. The HDDC models are fit using the default number of groups, whereas the $t$HDDC models are fit with $G = 1,\ldots,4$. It is important to note that the only HDDC models considered are ones with a monotonic likelihood.

\section{Applications}\label{sec:applic}
\subsection{Performance Assessment and Model Selection}
The data analyses will be treated as genuine clustering problems, where the true classifications are not known. Since we do have the true class labels, the adjusted Rand index \citep[ARI;][]{hubert1985comparing} will be used to assess class agreement between the true class labels and the predicted labels rendered by the clustering techniques. The ARI was introduced to correct the Rand Index \citep{rand71} for chance since the expected value of the Rand Index is greater than 0 for a random classification, making it hard to interpret. So, the ARI has expected value equal to `0', with a perfect classification being represented by a score of `1'. Formally, the ARI can be written as 
\begin{equation}
\text{ARI} = \frac{\text{number of pairwise agreements}}{\text{number of pairs}}.
\end{equation}
In all applications, the best fitting models will be chosen using the BIC.

\subsection{Simulation Studies}
\begin{figure}[!ht] 
	\includegraphics[height=3in]{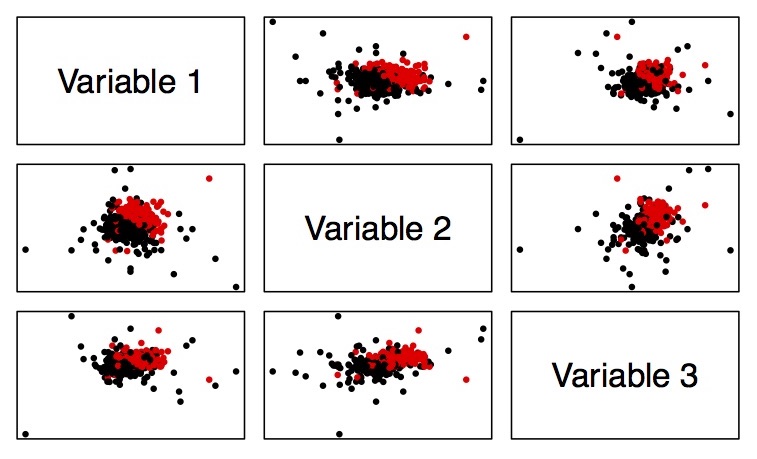}
	\centering
	\caption{Pairs plot of the first three dimensions of one multivariate-$t$ simulated data coloured by true groups.  \label{fig:sim_pairs}}
    \end{figure}

We use a simulation study to highlight the aforementioned drawback of the considered mixture of multivariate Gaussian distributions. Ten data sets were simulated from a two-component multivariate-$t$ distribution with $\nu_1=2$ and $\nu_2=3$. 
Figure~\ref{fig:sim_pairs} provides an illustration of the first three dimensions of one of the simulated data sets. In each component, observations are scattered from the mean, with many outliers on far ends of the clusters.
Table~\ref{tab:sim_t_res} gives the classification results for the tHDDC and HDDC models when fitted to the simulated data set. As expected, the tHDDC approach outperforms the HDDC approach, achieving a near perfect classification. The relatively small standard deviation reveals that the selected $t$HDDC models are consistently returning an exceptional classification performance, whereas the selected HDDC models are using extra components to account for the increased variation in the simulated data sets.
\begin{table}[!h]
\centering%
\caption{Mean and standard deviation of ARI values returned by the best fitting HDDC and $t$HDDC models found by the BIC for the simulated multivariate-$t$ data sets.\label{tab:sim_t_res}}
\begin{tabular*}{0.7\textwidth}{@{\extracolsep{\fill}}lcc}
\hline
&Mean of ARI &Standard Deviation of ARI\\
\hline
HDDC & 0.021&0.012\\
$t$HDDC &0.995&0.005 \\
\hline
\end{tabular*} 
\end{table} 

\subsection{Fisher's Irises}
In our first real data analysis, we consider Fisher's famous iris data set, which is available in the \textsf{R} package \texttt{datasets}. It is composed of four explanatory variables: sepal length, sepal width, petal length, and petal width, measured in centimetres. There are three species of the plant: Setosa, Versicolour and Virginica. Table~\ref{tab:iris_res} gives the classification results. 

\begin{table}[!h]
\centering%
\caption{Model decomposition, number of components, BIC and ARI values for the best fitting tHDDC and HDDC models found for the Iris data.\label{tab:iris_res}}
\begin{tabular*}{0.7\textwidth}{@{\extracolsep{\fill}}lrrrr}
\hline
& Model & $G$ & {BIC} & ARI\\
\hline
$t$HDDC & UUUCC & $3$ & $-646.327$ & 0.904 \\
HDDC & $\text{A}_{kj} \text{B}_k \text{Q}_k \text{D}$ & 3 & $-588.01$ & 0.868 \\
\hline
\end{tabular*} 
\end{table} 

Both the best fitting HDDC and tHDDC models return a very good classification of the irises, with the best fitting tHDDC model outperforming the corresponding HDDC model. Across the three selected components, the best fitting tHDDC model uses a varying number of eigenvalues and eigenvectors with a constant intrinsic dimension and degrees of freedom. In total, this model misclassifies only 5 irises (see Table~\ref{tab:iris_class}). 
\begin{table}[!ht]
\centering
\caption{A classification table showing the results for the selected three-component UUUCC $t$HDDC model for the iris data.\label{tab:iris_class}}
\begin{tabular*}{0.7\textwidth}{@{\extracolsep{\fill}}lccc}
\hline
& A&B&C\\
\hline
Setosa & 50 &0&0 \\
Versicolor & 0 &45&5\\
Virginica & 0&0&50 \\
\hline
\end{tabular*} 
\end{table}

\subsection{Italian Wines}
The Italian wines data set, available as \texttt{wine} in the \textsf{R} package \texttt{pgmm} \citep{mcnicholas2015pgmm}, is composed of 178 Italian wines on which 27 measurements are taken \citep{forina86}. The wines come from three different cultivars and are classified based on which one they come from: Barolo, Grignolino and Barbera. The $t$HDDC models are fit using $G=1,\ldots,5$,  since the BIC will select a four-component mixture model. Table~\ref{tab:wines_res} gives the classification results. Note: when fitting only three-component HDDC and tHDDC models, the selected $A_jBQD$ and GCCCC models gave the same classification result (ARI = 0.933).

\begin{table}[!ht]
\centering%
\caption{Model decomposition, number of components, BIC and ARI values for the best fitting HDDC and$t$ HDDC models when fitted for $g=1,\ldots,10$  and $g=1,\ldots,5$ components for the Italian wines data.\label{tab:wines_res}}
\begin{tabular*}{0.7\textwidth}{@{\extracolsep{\fill}}lrrrr}
\hline
& Model & $G$ & {BIC} & ARI\\
\hline
HDDC & $\text{A}_{j} \text{B} \text{Q} \text{D}$ & $8$ & $-12,071.63$ & 0.658\\
$t$HDDC & GCCCC & $4$ & $-11,965.26$ &  0.758\\
\hline
\end{tabular*} 
\end{table} 

We can see that the best fitting HDDC model overfits by selecting a model with eight components. Although the best $t$HDDC model did not have three groups, the four group solution gives a superior classification performance (See Table~\ref{tab:wines_res}). 

\subsection{Martian Surface}
\begin{figure*}[!ht]
\centering
\includegraphics[height=2.4in,width=0.95\textwidth]{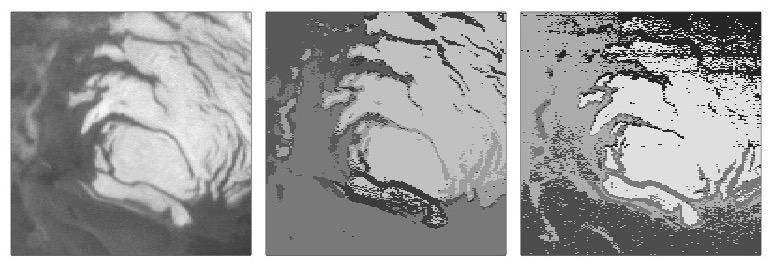}
\caption{Image based on spectral data collected by the OMEGA instrument (left), image based on classes predicted by $t$HDDC (middle) and image based on classes predicted by HDDC (right).
\label{fig:3way}}
\end{figure*}

This data set was retrieved by the OMEGA instrument \citep[Mars Express, ESA;][]{bibring2004omega}. The OMEGA instrument is used for characterization of the Martian surface based on physical and chemical composition. This can include classes of silicates, hydrated minerals, ices and more. The data used is based on one $300 \times 128$ raw image. It contains 255 variables on 38,400 observations. 
With a physical model, eight groups were found and for the purpose of this analysis, these will be treated as true groups; however, the best determination of model performance here is based on efficacy for image reconstruction. For $G=8$ components, $t$HDDC does a little better than HDDC; however, neither performs well (Table~\ref{tab:mars8_res}).
\begin{table}[!t]
\centering%
\caption{Model decomposition, BIC and ARI values for the selected eight-component HDDC and $t$HDDC models for the Martian surface data.\label{tab:mars8_res}}
\begin{tabular*}{0.7\textwidth}{@{\extracolsep{\fill}}lrrrr}
\hline
&  $G$ & Model & {BIC} & ARI\\
\hline
HDDC & $8$ & $A B_k Q_k D_k$ & $62,460,591$ & 0.319\\
$t$HDDC & $8$ & CCCCC & $64,249,591$ & 0.351\\
\hline
\end{tabular*} 
\end{table}

Although the physical model suggests eight groups, experts in the field are interested in exploring a five group solution \citep{bouveyron2007high}. Both HDDC and $t$HDDC models are applied to this data with $G=5$, and the selected $t$HDDC model recovers the clusters better than the selected HDDC model (see Table~\ref{tab:mars5_res}). 

\begin{table}[!ht]
\centering%
\caption{Model decomposition, BIC and ARI values for the selected five-component HDDC and $t$HDDC models for the Martian surface data. \label{tab:mars5_res}}
\begin{tabular*}{0.7\textwidth}{@{\extracolsep{\fill}}lrrrr}
\hline
&$G$&Best Model&{BIC}& ARI\\
\hline
HDDC &$5$&$A_{kj}B_kQ_kD_k$& $61,956,344 $&0.472\\
$t$HDDC&$5$&UCUUC & $70,120,085$&0.645\\
\hline
\end{tabular*} 
\end{table} 

In Table~\ref{tab:mars_versus}, we can see that the classification results returned by the selected $t$HDDC and HDDC models are quite different. 
\begin{table}[ht]
\centering
\caption{A classification table comparing the best fitting five-component $t$HDDC and HDDC models for the Martian surface data. \label{tab:mars_versus}}
\begin{tabular*}{0.7\textwidth}{@{\extracolsep{\fill}}llccccc}
\hline
&&\multicolumn{5}{c}{HDDC}\\
\cline{3-7}
&& A&B&C&D&E\\
\hline
&1& 10744 &22&0&0&2973\\
&2& 1019 &1598&0&99&785\\
$t$HDDC&3& 0 &39&7807&4372&4\\
&4& 2 &319&1&4871&716\\
&5& 605 &914&4&283&1223\\
\hline
\end{tabular*} 
\end{table} 

Furthermore, comparing the recovered image based on the predicted classes to the original image (see Figure~\ref{fig:3way}), the utility of the model becomes clear, i.e., the physical details are generally recovered very well.


\section{Discussion}\label{sec:discuss}
A new family of multivariate-$t$ mixture models has been proposed. The $t$HDDC approach is an extension of the HDDC approach that incorporates the multivariate-$t$ distribution, allowing for a more robust clustering scheme.  A total of 28 models have been developed and the need for these models was shown through a simulation study which demonstrated their flexibility in recognizing outliers. The models were tested on both simulated and real data sets and show superior results when compared to the HDDC family. In particular, the results on the high-dimensional Martian surface data show that image recovery can be greatly improved. Overall, the added degrees of freedom parameter allows for more flexible clusters and a more flexible modelling structure than HDDC. 
In future work, this method can be extended to include skewed mixture models. Examples include the mixture of multivariate skew-$t$ distributions \citep{lin10,murray14b,murray17}, the mixture of shifted asymmetric Laplace distributions \citep{franczak2014mixtures}, the mixture of variance-gamma distributions \citep{smcnicholas17}, and the mixture of generalized hyperbolic distributions \citep{browne15}.

\section*{Acknowledgements}
This work was supported by a Discovery Grants from the Natural Sciences and Engineering Research Council of Canada (Franczak, McNicholas) and the Canada Research Chairs Program (McNicholas).

{\small\bibliographystyle{chicago}
\bibliography{mybib}}

\end{document}